# Band degeneration and evolution in nonlinear triatomic chain superlattices

Chen Gong[1], Xin Fang,[2,1][*] and Li Cheng[1][†]

[1]*Department of Mechanical Engineering, Hong Kong Polytechnic University, Hung Hom, Kowloon, 999077, Hong Kong*
[2]*Laboratory of Science and Technology on Integrated Logistics Support, College of Intelligent Science and Technology, National University of Defense Technology, Changsha, Hunan 410073, China*

**Abastract:** Nonlinear superlattices exhibit unique features allowing for wave manipulations. Despite the increasing attention received, the underlying physical mechanisms and the evolution process of the band structures and bandgaps in strongly nonlinear superlattices remain unclear. Here we establish and examine strongly nonlinear superlattice models (three triatomic models) to show the evolution process of typical nonlinear band structures based on analytical and numerical approaches. We find that the strongly nonlinear superlattices present particular band degeneration and bifurcation, accompanied with the vibration mode transfer in their unit cells. The evolution processes and the physical mechanisms of the band degeneration in different models are clarified with the consideration of the mode transfer. The observed degeneration may occur as the shifting, bifurcating, shortening, merging or disappearing of dispersion curves, all depending on the arrangement of the coupled nonlinear elements. Meanwhile, the dimension of the unit cell reduces, alongside changes in the frequency range and mechanisms (Bragg and local resonance) of the bandgaps. These findings answer some foundamental questions peritinent to the study of nonlinear periodic structures, nonlinear crystals and nonlinear metamaterials, which are of interest to the broad community of physics.

## I. INTRODUCTION

Band structures are essential features of periodic structures including conventional crystals and more advanced superlattices [1-3]. Metamaterials are superlattice structures featuring exceptional subwavelength functionalities which are unusual to nature materials such as perfect absorption, negative refraction, asymmetric transmission and phase modulation etc. [4-10]. Despite the extensive attention received, most studies on superlattices and metamaterials are limited to linear systems. Based on Bloch theorem, band structures of various types of linear superlattices are well understood and tactically structured to achieve nonclassical wave manipulations exemplified by topological insulators. Bandgaps can also entail efficient suppression of elastic waves and vibrations, a salient feature that can be obtained through manipulating band gaps and the subsequent wave propagation by tuning material and structural parameters.

Nonlinear superlattices can exhibit behaviors that are inaccessible and potentially superior to their linear counterparts, such as harmonic generation [11,12], phase matching [13,14], nonlinear resonance shifting [15-17], dispersion modulation [18,19], chaos [11,20,21], breathers [22-24] and solitons [25,26] etc.. Recently, strongly nonlinear elastic/acoustic metamaterials (NAMs) consisting of periodic nonlinear local resonators were shown to offer ultralow and ultrabroad-band vibration suppression arising from the chaotic band effects [27] and the bridging coupling of bandgaps [21] among other salient features specific to nonlinear superlattice. The Bragg and locally resonant bandgaps in a NAM are amplitude-dependent [28-30]. This property has been extensively investigated using the perturbation method [31,32], harmonic balance method [19], homotopy analysis method [20] and equivalent method [30,33-35]. It was theoretically and experimentally demonstrated that band-gap effect in a strongly NAM can be adaptively broadened as the propagation distance/time increases [35] leading to broadband acoustic limiting. With increasing amplitude, bandgaps in a NAM show complexes variation such as shifting, switching and eventually mutual coupling [29,36]. Wave transmission within nonlinear bandgaps also features multiple stability behavior. Excessive nonlinearity may even lead to the disappearance of some bandgaps [20]. However, the underlying physical mechanisms and the evolution process of these peculiar bandgap phenomena remain largely unclear. For example, as a result of the bandgap shifting under strongly nonlinearity, the band structure reflected by the dispersion curves changes; but the underpinning reasons, the associated evolution process as well as its influences on the wave propagation remains unknown. This is seen as one of the bottlenecking problems which hinder the exploration of the NAM-specific properties as well as the applications of strongly nonlinear superlattices.

---

[*] xinfangdr@sina.com
[†] li.cheng@polyu.edu.hk

In this paper, we establish three one-dimensional superlattice models to elucidate the evolution process of typical nonlinear band structures and explore the underlying physical mechanisms. In section 2, we introduce the models alongside the analytical method for band structure analysis. In section 3, we take a typical case to demonstrate our methods. In section 4, based on a typical model, we report and scrutinize the band degeneration and merging phenomenon, and analyze the evolution of the band structure and the associated vibration modes. In addition, we explore the adaptive broadening band-gap effect and validate the applicability of Bloch theorem. To show the generality of the band degeneration, we consider two other models in section 5. In addition to the discovery of other types of band degeneration, we show the property of dimensionality reduction and establish the common features and the relation in these three nonlinear models.

## II. SUPERLATTICE MODEL AND ANALYSIS METHODS

A nonlinear triatomic chain with two local resonators offers two nonlinear resonant bandgaps and a Bragg bandgap that may shift, merge, switch and couple. Therefore, a triatomic chain is a versatile superlattice model that can showcase typical wave dynamics in nonlinear superlattices. Here, we establish an infinite model to investigate the dispersion properties and wave propagation of this strongly nonlinear superlattice model. As shown in Fig. 1, a unit cell contains three oscillators with respective masses $m_0$, $m_1$ and $m_2$ interconnected by different springs. As such, nonlinearity can appear between $m_1$-$m_2$, or $m_0$-$m_2$, or both, which will alter the wave propagation. In the first step, we take the case with nonlinearity between $m_1$ and $m_2$ as an example to demonstrate the method and show typical phenomena, before considering other cases. More specifically, cubic stiffness nonlinearity $k_1 p + k_N p^3$ is considered, with $k_1$ and $k_N$ denoting the linear and nonlinear stiffness coefficients, respectively.

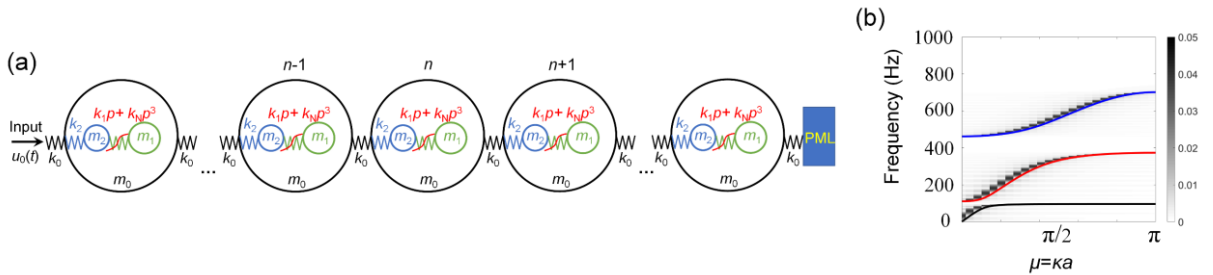

FIG. 1. Superlattice Model. (a) Schematic of the periodic triatomic chain. (b) Linear dispersion curves of the model calculated by analytical and numerical methods ($k_N=0$). The solid curves are analytical results, and the gray shading regions are numerical results.

In this case, the neighboring oscillators $m_0$ in the primary chain are coupled through linear stiffness $k_0$, and $m_2$ couples to $m_0$ through a linear stiffness $k_2$. Damping is not considered. The displacements of $m_0$, $m_1$ and $m_2$ are $u_n$ and $y_n$ and $z_n$, respectively, and $p_n = y_n - z_n$. Thus, the equations of motion for the $n^{th}$ unit cell write:

$$\begin{cases} m_0 \ddot{u}_n = k_0(u_{n+1} + u_{n-1} - 2u_n) + k_2(z_n - u_n) \\ m_1 \ddot{y}_n = -k_1 p_n - k_N p_n^3 \\ m_2 \ddot{z}_n = -k_2(z_n - u_n) + k_1 p_n + k_N p_n^3 \end{cases} \quad (1)$$

Mindful of the possible deficiencies of the Bloch-Floquet theorem in dealing with nonlinear systems [35], we still use it to solve the above system to get the dispersion curves and then clarify its limitations hereafter. According to the Bloch-Floquet theorem, $u_{n+1}=u_n e^{-i\kappa a}$, $u_{n-1}=u_n e^{i\kappa a}$, $\kappa a=\mu \in [0, \pi]$, the motion equation of the $n^{th}$ unit cell can be written as:

$$\mathbf{M\ddot{X}} + \mathbf{KX} + \mathbf{K_N X}^3 = 0 \tag{2}$$

where

$$\mathbf{M} = \begin{bmatrix} m_0 & 0 & 0 \\ 0 & m_1 & m_1 \\ 0 & 0 & m_2 \end{bmatrix}, \mathbf{K} = \begin{bmatrix} k_2 - k_0(e^{-ika} + e^{ika} - 2) & 0 & -k_2 \\ 0 & k_1 & 0 \\ -k_2 & -k_1 & k_2 \end{bmatrix},$$

$$\mathbf{K_N} = \begin{bmatrix} 0 & 0 & 0 \\ 0 & k_N & 0 \\ 0 & -k_N & 0 \end{bmatrix} \text{ and } \mathbf{X} = \begin{bmatrix} u_n \\ p_n \\ z_n \end{bmatrix}$$

Natural frequencies of individual oscillators before they are coupled together are denoted by $\omega_i = \sqrt{k_i/m_i} = 2\pi f_i$, $i$=0, 1, 2. Taken from a recent experimental configuration [35], parameters used in the simulation are: $a$=1, $m_0$=5.8, $m_1$=2.1, $m_2$=2 g; $f_0$=322, $f_1$=100, and $f_2$=390.6 Hz. We took $k_N$=1 × $10^{13}$ N/m³ to show the nonlinear phenomena.

We adopt the first-order harmonic balance method to solve Eq. (2) by assuming the solution as

$$\mathbf{X} = \mathbf{A} \sin(\omega t) \tag{3}$$

in which $\mathbf{A}$=[$A_0, A_{12}, A_2$]$^T$ with $A_0$, $A_{12}$ and $A_2$ standing for the amplitudes of $u_n$, $p_n$ and $z_n$, respectively; $\omega$=2$\pi f$. Substituting Eq. (3) into (2) and balancing the coefficients of sin($\omega t$) give:

$$[K - \omega^2 M]\mathbf{A} + \frac{3}{4} K_N \mathbf{A}^3 = 0 \tag{4}$$

By specifying $A_0$ for a given wave number $\mu=\kappa a \in [0, \pi]$, the eigenfrequency $\omega$ and the eigenvector [$A_0, A_{12}, A_2$]$^T$ can be obtained from Eq. (4). $A_1$ is the amplitude of $y_n$, and $A_1=A_{12} + A_2$, thus, the normalized eigenvector [$A_0, A_1, A_2$]$^T/A_0$ represents the corresponding vibration modes of the superlattice unit cell.

Meanwhile, we adopt a numerical approach to calculate the dispersion curves. In the simulation, we establish a model consisting of 2000 triatomic cells. An optimized perfect matching layer is added to the right end of the chain to suppress possible wave reflection. A displacement excitation, $u_0(t)$, is imposed from the left end of the chain. $u_0(t)$ is a chirp harmonic wave whose frequency increases from 0 to 1000 Hz within 1 s. For the metacells far enough from the excitation, it takes more than 1 s for the wave to reach them. Thus, the duration of the response is set at 2 s (i.e., the simulation time) to get a complete response, which is enough for the first 500 metacells to respond (in the present case, we just involve the calculation of the first 200 metacell). Once the time-domain response $u(n, t)$ for the $n^{th}$ cell is extracted, we conduct a 2D Fast Fourier Transform (2DFFT) to obtain the 2D frequency spectrum:

$$V(\mu, f) = \sum_{n=2}^{N-1} \int_0^T u(n,t)e^{-j(\mu n + \omega t)} dt \tag{5}$$

where $N$ denotes the number of cells used in the 2DFFT, and $T$=2 s denoting the time interval for the integral. While transforming the time domain signal into the frequency domain by 2DFFT, the space variable $n$ is also transformed into the wave number $\mu$. Therefore, the spectrum $V(\mu, f)$ represents the dispersion curves. A larger $N$ offers a higher resolution of the wave vector $\mu$ in $V(\mu, f)$. However, $N$ cannot be excessively large due to the self-adaptive band structure (see Appendix). The 2DFFT is conducted using the signals of the first 50 cells ($N$=50) counting from the excitation position, thus presenting the dispersion curves for a given excitation level $A_0$.

To validate the method, we calculate the dispersion curves of the corresponding linear superlattice model (with $k_N$=0). As shown in Fig. 1(b), the analytical and numerical dispersion curves agree well with each other. This particular configuration entails a Bragg bandgap (from 700 Hz on) and two locally bandgaps LR1 (96.1, 110.4) Hz and LR2 (373.4, 462) Hz.

## III. SALIENT PROPERTIES OF NONLINEAR BAND STRUCTURE

In this section, we explore typical dispersion properties of the nonlinear superlattice model with $A_0$=5 μm, as shown in Fig. 2. The complex frequency solutions, $f=f_R + if_I$, are obtained for any given wave vector $\mu$, where $f_r$ and $f_I$ denote the real and imaginary parts, respectively. As shown in Figs. 2 (a) and (b), the harmonic balance solution presents six dispersion curves, instead of three in the linear model. Several interesting phenomena are noteworthy. All solutions on curves 1 and 4 are real numbers ($f_I$=0), which shift upwards relative to the linear curves, and the part in $\mu \in [0, 0.5]$ on curve 4 is also uplifted. Curves 2 and 3 correspond to the conjugate solutions with very large $f_I$. Interestingly, curves 5 and 6 feature a bifurcation at the specific wave vector $\mu_{Bif}$ ($\mu_{Bif}$=0.56 for the present case with $A_0$=5 μm).

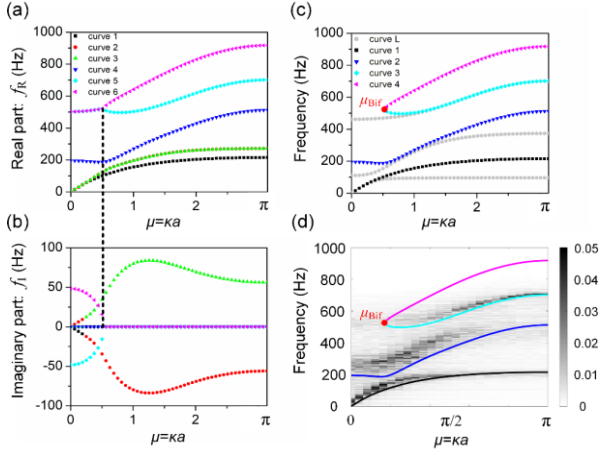

FIG. 2. Dispersion curves of nonlinear the superlattice model with $A_0$=5 μm. (a) and (b) The real part $f_R$ and imaginary part $f_I$ of the curves. There are 6 curves. The corresponding curves in the upper and lower panels use the same type of lines. (c) Curves only with real numbers (i.e., $f_I$=0). Herein, the gray curves are the linear results. (d) Numerical and analytical dispersion curves.

In $\mu \in [0, \mu_{Bif}]$, solutions on curves 5 and 6 are conjugate with large $f_I$; while in $\mu \in [\mu_{Bif}, \pi]$, they bifurcate into two branches with zero imaginary part $f_I$=0. This means the two dispersion curves degenerate at $\mu_{Bif}$, a degeneration induced by the nonlinearity. Expressing the wave in the nonlinear lattice as $U(x,t)=A\exp(\omega t - kx)$, any solution component with large $\omega_I \neq 0$ corresponds to the wave with rapid attenuation. Therefore, we can remove these solutions from the band structure and finally obtain four dispersion curves shown in Fig. 2(c). Now they are re-numbered as curves 1 to 4. As shown in Fig. 2(c), the numerical solution is consistent with the analytical solution. Some differences appear due to the wave amplitude which varies in the chain (i.e., in the analytical solution, all amplitudes in $n < 50$ are set to $A_0$, but actually it is not a constant across the units). Importantly, the numerical curve 3 is also noncomplete—the part in $[0, \mu_{Bif}]$ disappears, as the analytical curve does. This confirms the degeneration of the nonlinear band structure. Meanwhile, the numerical results fail to predict the curve corresponding to the analytically curve 4. The following study will clarify its observation.

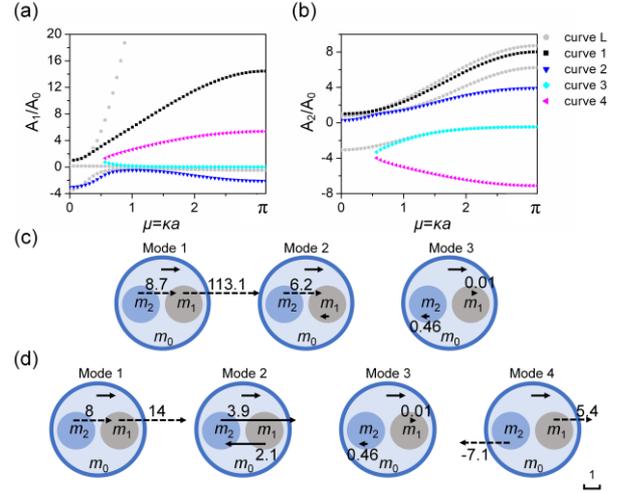

FIG. 3. Normalized vectors and vibration modes of the linear and nonlinear superlattice. $A_0$=5 μm. (a) and (b) Normalized vectors $A_1/A_0$ and $A_2/A_0$ with respect to the wave number $\mu$, respectively. Herein, curve L and curves 1-4 denote the linear and nonlinear cases, respectively. Colors used in (a) and (b) correspond to the curves with the same colors in Fig. 2(c). (c) and (d) Vibration modes for linear and nonlinear superlattice at $\mu=\pi$, respectively. Mode $j$ corresponds to curve $j$ in (a) and (b). The length and the direction of arrows denote the generalized length $A_i/A_0$ and the corresponding phase (positive or negative), respectively; some numbers (value=$A_i/A_0$,) are labeled for very long arrows.

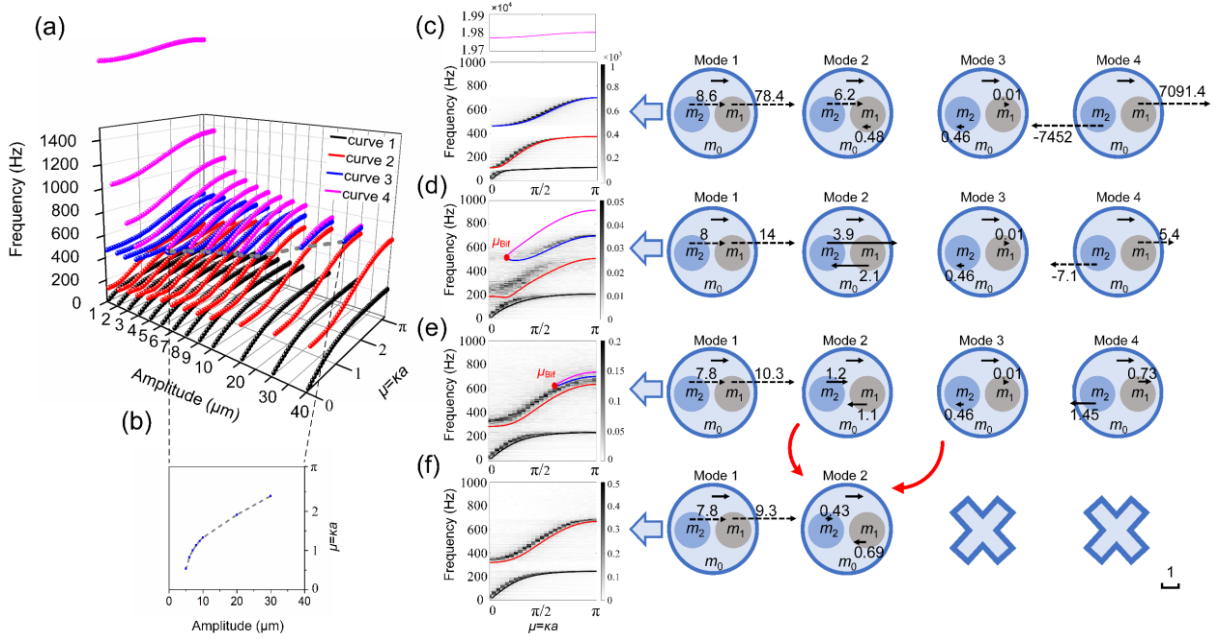

FIG. 4. Evolution of dispersion curves and vibration modes with increasing incident amplitude amplitude $A_0$. (a) Overall view of analytical dispersion curves varying with increasing amplitude $A_0$. (b) Trace of bifurcation in (a). (c, d, e, f) Four typical cases showing their analytical and numerical dispersion curves and vibration modes; (c) Case-1: weak nonlinearity ($A_0$=0.1 μm), (d) Case-2: moderate nonlinearity ($A_0$=5 μm), (e) Case-3: strong nonlinearity ($A_0$=30 μm), (f) Case-4: stronger nonlinearity ($A_0$=40 μm).

On the other hand, the normalized vector $[A_0, A_1, A_2]^T/A_0$ from the analytical solution represents a vibration mode of the cell. The normalized displacements of $m_1$ ($A_1/A_0$) and $m_2$ ($A_2/A_0$) are shown in Figs. 3(a) and (b), respectively. The vibration modes at $\mu=\pi$ include the phase between different oscillators and their respective vibration displacement amplitudes. They are important metrics of local resonances and scattering, showing the essence of the bandgaps. As shown in Fig. 3(c, d), three circles represent the oscillators $m_0$, $m_1$, and $m_2$; and vectors represent their vibration modes. Similarly, the direction and the length of a vector denote the phase and displacement, respectively.

For the linear superlattice in Fig. 2(c) and Fig. 3(c), mode 1 is the local resonance of $m_1$; mode 2 is the local resonance of both $m_1$ and $m_2$ (vibrating in phase), respectively; in mode 3, the displacements of $m_1$ and $m_2$ are tiny relative to that of $m_0$, which belongs to the Bragg scattering of $m_0$ at the periodic lattice interface. The observed resonance and scattering inform on the well-known mechanisms of locally resonant bandgap and Bragg bandgap in linear superlattice [6,7]. For the nonlinear superlattice in Fig. 3(d), modes 1 and 2 still show typical local resonances, but $A_1/A_0$ in mode 1 and $A_2/A_0$ in mode 2 decrease greatly relative to the corresponding linear modes and curves (see Fig. 3(a, b)). $A_1/A_2$ in modes 1 and 2 change from 13 and 0.07 in the linear superlattice, further to 1.75 and 0.54 in the nonlinear superlattice, which suggests that both $m_1$ and $m_2$ take part in the local resonance in the nonlinear superlattice. Mode 3 remains nearly unchanged at $\mu=\pi$ for $A_0$=5 μm. Moreover, the nonlinear superlattice possesses an extra mode 4 owing to an extra dispersion curve 4 in Fig. 2(c). In this mode, $m_0$ and $m_2$ vibrate with opposite phase, $|A_2| \gg A_0$ but $|A_2| \gg |A_1|$, which indicates an inverse resonance of $m_2$ (different from the resonance of modes 1 and 2). More features of this mode will be scrutinized in the following. These phenomena show the band degeneration and mode transition in nonlinear superlattice. Next, we will study their evolution process and the underlying mechanisms.

## IV. EVOLUTIONARY PROCESS OF THE NONLINEAR BAND STRUCTURE

Figure 4 shows the variation process of the dispersion curves and vibration modes when $A_0$ increases. Interestingly, the harmonic balance method predicts 4 dispersion curves for weakly nonlinearity in mathematics.

With increasing $A_0$, curves 3 and 4 (the blue and pink ones) disconnect for $A_0 \to 0$, connect at $\mu=0$ for $A_0=4.4$ μm, and then degenerate (becomes incomplete) and bifurcate at $0 < \mu_{Bif} < \pi$. At last, curves 3 and 4 disappear under very strong nonlinearity: only two curves are left behind. Moreover, the behaviors manifested by curves 1~4 are accurately confirmed by numerical results.

We take four typical cases to explain the aforementioned degeneration and bifurcation process, as shown in Fig. 4(c-f). When nonlinearity is negligibly weak (the case $A_0=0.1$ μm in Fig. 4(c)), four dispersion curves are present mathematically. Curves 1~3 are all below 700 Hz while curve 4 appears in 19810~19790 Hz at $A_0=0.1$ μm. Except for curve 4, all other analytical curves agree with those from simulations that are identical with the linear band structure. Different from modes 1 and 2, curve 4 represents the reverse resonances of $m_1$ and $m_2$, in which $A_2/A_0$ and $A_1/A_0$ abnormally reach $-7452$ and $7091$ for $A_0=0.1$ μm, respectively. As the nonlinear force is $k_1(A_1-A_2) + k_N(A_1-A_2)^3$, this means that this nonlinearity should become extremely strong, contradicting to the small input $A_0=0.1$ μm. Our simulation method by inputting wave energy $u_0(t)$ from the left end of the chain fails to get this curve that matches the analytical result, from weak to moderate nonlinearity range in Fig. 4(d). Presenting this abnormal mode in simulation may requires accurately controlling the phase and displacement of all oscillators. In this paper, we will focus on curves 1~3 to clarify other phenomena and mechanisms.

With further enhanced nonlinearity by increasing $A_0$, as shown in Fig. 4(a), curve 1 (the black one) shifts upwards from (0, 173.3) to (0, 242.6) Hz, and curve 2 (the red one) from (118.6, 379.8) to (313.4, 661.3) Hz. This leads to the modulation of bandgap 1 between curves 1 and 2: it first becomes blind at $A_0 \approx 1$ μm and then opens again when $A_0 > 8$ μm. The vibration modes tell the mechanism. By increasing $A_0$ from zero to 40 μm, $A_1/A_2$ decreases from 9.1 (in Fig. 4) to 1.2, which suggests the dominant mechanism for bandgap 1 changes from "the local resonance of $m_1$" to "the synchronous local resonance of $m_1$ and $m_2$".

The shifting of curve 2 is highly relevant to the degeneration of curves 3 and 4 in this process. The bifurcation point $\mu_{Bif}$ of curves 3 and 4 increases from 0 to π as $A_0$ increases from 4.4 to 37.8 μm (see Fig. 4(b)). Meanwhile, curves 3 and 4 shorten, merge and disappear at last. However, the cutoff frequency of curve 3 (when it exists) remains almost unchanged. The merging takes place as a result of the downward shifting of curve 4. Moreover, the cutoff frequency of curve 2 increases just right to the cutoff frequency of curve 3. "Disappearance of curve 3" occurs at the same $A_0$ as "the arriving of curve 2". This alludes to an interesting suspicion: curve 2 takes over the role of curve 3 when enhancing nonlinearity from weak to extremely strong state. Sure enough, an analysis on the variation of modes 2 and 3 would confirm this. As shown in Fig. 4(c-f), when $A_0$ increases from zero to 30 μm, the phases of $m_1$ and $m_2$ remain opposite, $|A_2/A_1|$ changes slightly, but $A_2/A_0$ decreases from 6.2 to 1.2, and $A_2/A_0$ finally decreases to 0.43 at $A_0=40$ μm. In the whole process, mode 3 nearly remains invariant. The only difference between mode 2 and mode 3 at $A_0=40$ μm is the sign of $A_2/A_0$. As $A_2/A_0 \to 0$, the positive or negative signs negligibly influence the wave dynamics. Therefore, same with mode 3, mode 2 at $A_0=40$ μm represents the Bragg scattering. This means all modes 2, 3 and 4 merge and behave as Bragg scattering under very strong nonlinearity. Thus, although the shape of curve 2 is not significantly altered, the mechanism for the responses changes from the "local resonance of $m_1$ and $m_2$" to "scattering between $m_0$". In this process, the second locally resonant bandgap between curves 2 and 3 becomes blind at $A_0 \approx 4.3$ μm but it does not appear again.

Overall, with the nonlinearity level increased, three bandgaps retreat and degenerate into two bandgaps owing to the merging of modes 2, 3 and 4. In other words, the nonlinear metacell transforms a 3DoF unit cell to a 2DoF one under strong nonlinearity. The observed dimensionality reduction with enhanced nonlinearity is unique and specific to strongly nonlinear superlattice model. This property will be explored in other strongly nonlinear systems hereafter so that some universal relations be extracted and established in section 5.

## V. EVOLUTION OF THE BAND STRUCTURE IN OTHER TYPICAL MODELs.

Here, we consider two other typical models to show the generality of the band generation and degeneration phenomena observed above. In the first model, the position of the nonlinear component is changed and placed between $m_2$ and $m_0$, while the connection between $m_1$ and $m_2$ becomes linear, as shown in Fig. 5(a). In the second model, both connections are nonlinear to represent a complex nonlinear coupling case, as shown in Fig. 7.

### A. Nonlinearity with exchanged position

Denoting the displacements of $m_0$, $m_1$ and $m_2$ by $u_n$, $y_n$ and $z_n$, respectively and assuming $q_n = z_n - u_n$, the equations of motion in a metacell shown in Fig. 5(a) write:

$$\begin{cases} m_0 \ddot{u}_n = k_0(u_{n+1} + u_{n-1} - 2u_n) + k_2 q_n + k_2 q_n^3 \\ m_1 \ddot{y}_n = -k_1(y_n - z_n) \\ m_2 \ddot{z}_n = k_1(y_n - z_n) - k_2 q_n - k_N q_n^3 \end{cases} \quad (6)$$

Following the same analytical and numerical procedure as detailed in section 3, we calculate the corresponding dispersion curves and vibration modes. Fig. 5(b) shows the variation of the dispersion curves and that of the vibration modes when increasing the incident amplitude $A_0$. Unlike the band structure in Fig. 4(a), only three dispersion curves appear irrespective of the amplitude value $A_0$. When increasing $A_0$, the portion $\mu \in [0, \mu_{bend}]$ on curve 3 bends upwards, with the point $\mu_{bend}$ shifting from 0 to $\pi$. At last, the entire curve 3 shifts upwards to a high frequency under very strong nonlinearity. Moreover, the variation trends manifested by curves 1~3 are also confirmed by numerical results.

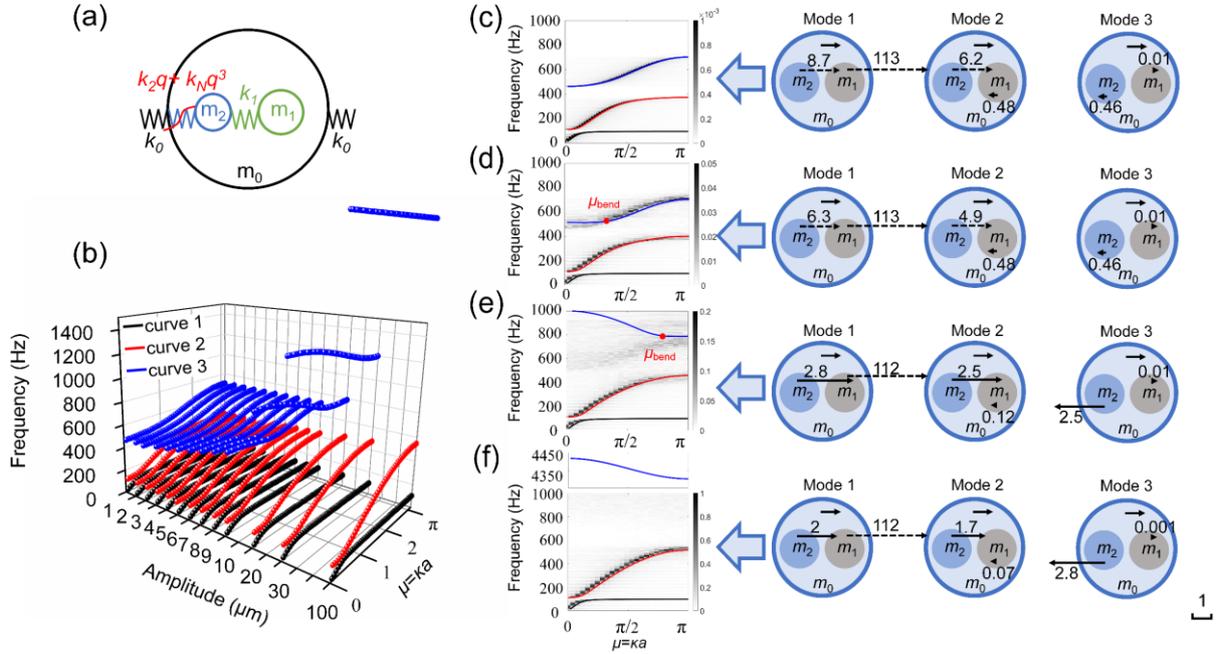

FIG. 5. Evolution of dispersion curves and vibration modes with increasing incident amplitude $A_0$ when the position of nonlinearity is exchanged. (a) Triatomic metacell. (b) Analytical dispersion curves with respect to increasing amplitude $A_0$. (c, d, e, f) Four typical cases showing their analytical and numerical dispersion curves and vibration modes; (c) Case-1: weak nonlinearity ($A_0$=0.1 μm), (d) Case-2: moderate nonlinearity ($A_0$=5 μm), (e) Case-3: strong nonlinearity ($A_0$=30 μm), (f) Case-4: very strong nonlinearity ($A_0$=100 μm).

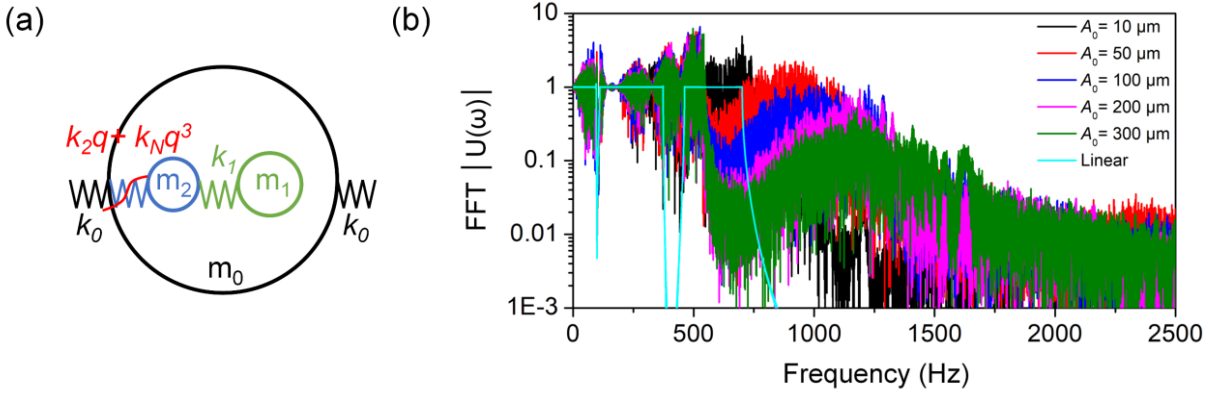

FIG. 6. Normalized frequency spectra of the response of the 5th metacell under a chirp wave excitation in the range of 0-2500 Hz. (a) Triatomic metacell. (b) The frequency spectra were obtained by the fast Fourier transform (FFT).

We also examine four typical cases to explain the aforementioned bending and shifting phenomena in Fig. 5(c-f). Unlike the nonlinear system discussed in section 4.1, no degeneration of curves is noticed. Curve 1 (the black one) shifts upwards from (0, 97.2) to (0, 99.1) Hz and curve 2 (the red one) from (110.8, 404.2) to (111.2, 515.3) Hz. The observed changes are consistent with those obtained from simulations as a result of the nonlinearity enhancement. Mode 1 always behaves as the local resonance of $m_1$ although $A_2/A_0$ decreases slightly. By contrast, $A_2/A_0$ in mode 2 decreases from 6.2 at $A_0$=0.1 μm to 1.7 at $A_0$=100 μm ($A_2/A_0 \rightarrow 1$). This happens because the stiffness between $m_2$ and $m_0$ expressed by $k_1+3k_N(A_2 - A_0)^2$ becomes much larger than its linear counterpart $k_1$, and thus $m_2$ is "fixed" onto $m_0$. As a result, mode 2 transforms from the local resonance of $m_2$ to the Bragg scattering between the "merged" mass $m_0 + m_2$.

Mode 3 for small input represents the Bragg scattering, but $A_2/A_0$ increases from 0.46 to 2.8 under strong nonlinearity. Therefore, curve 3 represents the high frequency resonance of $m_2$ under the strong nonlinearity. Meanwhile, curve 3 agrees well with the simulated one in the weak nonlinearity case. However, when the nonlinearity increases to the moderate case, the portion $\mu \in [0, \mu_{bend}]$ on curve 3 disappears in simulation. The frequency spectrum covering curve 3 at $A_0$=30 μm is continuous, which actually indicates the occurrence of chaotic responses [20]. When the nonlinearity becomes extremely strong, the entire curve 3 (not only the part in $\mu \in [0, \mu_{bend}]$) shifts to a high frequency range, although it is difficult to find such a curve in the 2DFFT simulation. Fortunately, we can still find this curve on the first 5 metacells close to the excitation, as shown in Fig.6. It is obvious that the third passband for curve 3 shifts from ~600 Hz to ~1250 Hz from linear state to the strongly nonlinear state with $A_0$=300 μm. Therefore, this curve does exist, but it appears only near the incident source region and its frequency range deviates from the analytical solution. A plausible reason is attributed to the self-adaptive band structure (see Appendix). We note that curve 3 under large input appears in the bandgap of its linear counterpart. As demonstrated earlier, the band structure of the nonlinear superlattice is self-adaptive. Due to the chaotic responses and harmonic generation, waves in the passband of nonlinear superlattice still undergo attenuations as the propagation distance increases. Due to the amplitude-dependent property, an attenuated amplitude leads to a different band structure. The linear bandgap finally appears when the amplitude becomes small. For example, although curve 3 shifts to 1250 Hz for $A_0$=300 μm near the excitation source, it rapidly attenuates as the propagation distance increases. Moreover, though the incident amplitudes $A_0$ are 50, 100, 200 and 300 μm, the mean amplitudes at the $5^{th}$ unit cell are much smaller, i.e. 23, 34, 54, and 90 μm. Thus, the numerical results deviate from the numerical results.

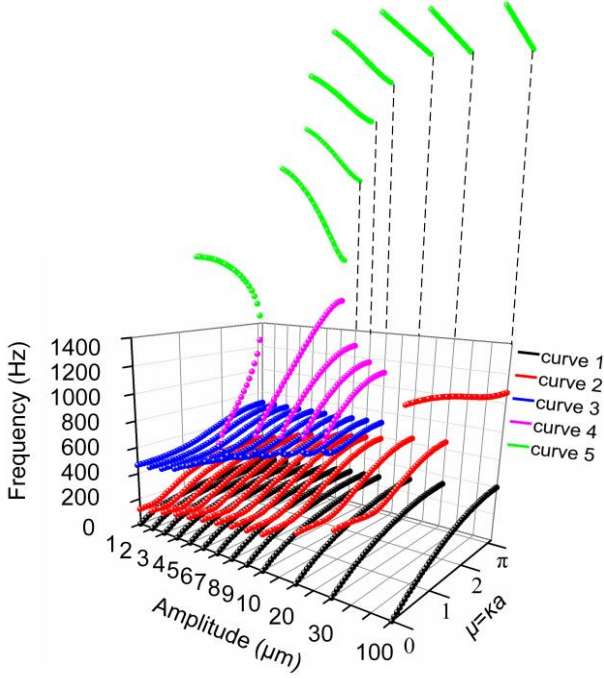

FIG. 7. Analytical dispersion curves with increasing amplitude $A_0$ in the complex nonlinear coupling case.

In short, although curve 3 does exist, its response disappears in an infinite chain, and the nonlinear system is converted from 3DoF to 2DoF systems with enhanced nonlinearity. Although the nonlinear systems in section 4.1 and section 5.1 both evolve from 3DoF to 2DoF systems when nonlinearity increases, they are different in principle. While the band degeneration in the former model undergoes shortening, merging of a dispersion curve but the first resonant bandgap is broadened; while the band degeneration in the latter model mainly features a curve shifting, response attenuation and a narrowing-down of the first resonant bandgap. Therefore, the dimensionality reduction in strongly nonlinear superlattice model with enhancing nonlinearity may take place in different ways.

### B. Complex nonlinear coupling

As demonstrated in sections 4.1 and 5.1, the nonlinear system with one nonlinear oscillator can evolve from a three-dimension to a two-dimension system with enhancing nonlinearity. One could surmise that if the two resonators all become nonlinear as Fig. 8(a), the metacell would then evolve from 3DoF to a 1DOF system with enhanced nonlinearity. This is investigated here. The displacements of $m_0$, $m_1$ and $m_2$ are $u_n$, $y_n$ and $z_n$, respectively, $p_n = y_n - z_n$, $q_n = z_n - u_n$, and the equations of motion of in a metacell read:

$$\begin{cases} m_0 \ddot{u}_n = k_0(u_{n+1} + u_{n-1} - 2u_n) + k_2 q_n + k_2 q_n^3 \\ m_1 \ddot{y}_n = -k_1 p_n - k_N p_n^3 \\ m_2 \ddot{z}_n = -k_2 q_n - k_N q_n^3 + k_1 p_n + k_N p_n^3 \end{cases} \quad (7)$$

Following the same analysis method in section 3, Fig. 7 shows the variation of the dispersion curves and vibration modes with respect to increasing incident amplitude $A_0$ in this complex nonlinear coupling case. This superlattice features both type of the band degeneration phenomena discussed in section 4.1 and 5.1. The harmonic balance method predicts 3 dispersion curves when nonlinearity is weak. Interestingly, curves 4 and 5 appear with increasing $A_0$ before bifurcating at $0 < \mu_{Bif1} < \pi$. At last, curves 4 and 5 disconnect at $\mu = \pi$ under moderate nonlinearity and shift downwards and upwards, respectively. Then, similar to the degeneration process in the first model, curves 3 and 4 degenerate and bifurcate at $0 < \mu_{Bif2} < \pi$. At last, curves 3 and 4 disappear under strong nonlinearity with only three curves left behind. When further increasing $A_0$, curve 2 shifts upwards, similar to that of curve 3 discussed in section 5.1. Moreover, the behaviors manifested by curves 1~4 are accurately confirmed by the numerical results.

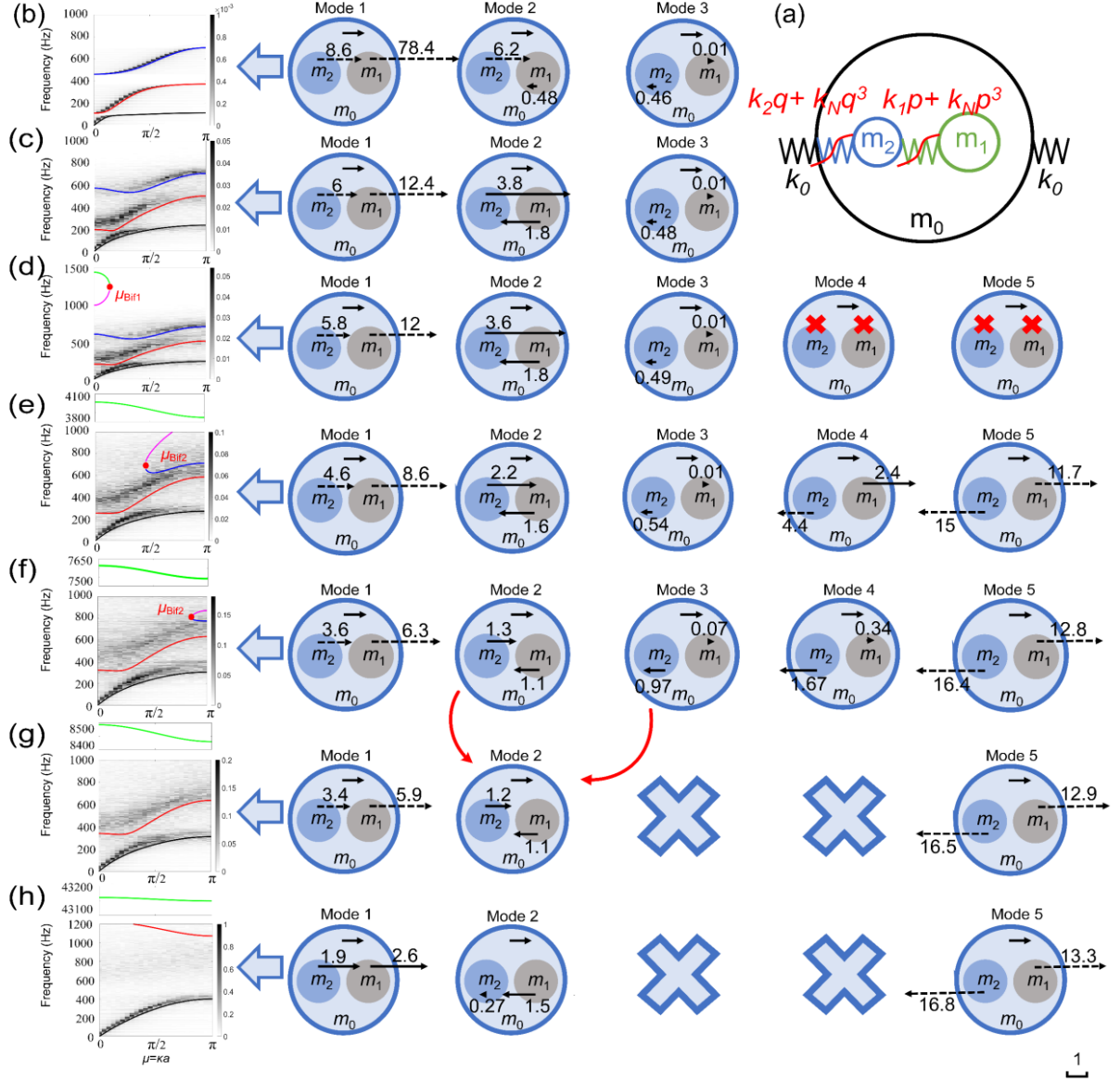

FIG. 8. Evolution of dispersion curves and vibration modes with increasing incident amplitude amplitude $A_0$ in the complex nonlinear coupling case. (a) Triatomic metacell. (b, c, d, e, f, e, f, g, h) Seven typical cases showing their analytical and numerical dispersion curves and vibration modes; (b) Case-1: weak nonlinearity ($A_0$=0.1 um), (c) Case-2: moderate nonlinearity 1 ($A_0$=5 um), (d) Case-3: moderate nonlinearity 2 ($A_0$=5.4 um), (e) Case-4: strong nonlinearity 1 ($A_0$=10 um), (f) Case-5: strong nonlinearity 2 ($A_0$=18 um), (g) Case-6: strong nonlinearity 3 ($A_0$=20 um) (h) Case-7: strong nonlinearity 4 ($A_0$=100 um).

We take seven typical cases to investigate the behavior of dispersion curves and the evolution of vibration modes in Fig. 8(b-h). When the nonlinearity is negligibly weak (the case $A_0$=0.1 μm in Fig. 8(c)), there exist three dispersion curves mathematically, and all analytical curves agree with those from the simulations, which are identical to the linear band structure. When nonlinearity evolves from moderate to strong (the cases $A_0$=5.4, 10 and 18 μm in Fig. 8(d-f)), complex changes take place. In short, curves 3 and 4 are highlighted by a "shortening, merging, and disappearance" process, similar to the first model in section 4.1. At $A_0$=7 um, curves 4 and 5 disconnect at $\mu=\pi$, but for $A_0 < 7$ um, curves 4 and 5 have no vibration mode at $\mu=\pi$ (i.e., they appear at $\mu$ =0 but cannot reach $\mu=\pi$).

Mode 4 degenerates from local resonance to Bragg scattering. Mode 5 shows that curve 5 always involve the reverse resonances of both $m_1$ and $m_2$, like the reverse resonance in Fig. 4, the simulation method also fails to present them. When the nonlinearity is extremely strong (the case with $A_0$=100 μm in Fig. 8(h)), curve 2 cannot be found in the dispersion curve, but it can be confirmed in the detailed analysis, like the curve 3 of the second model in section 5.1. We compare the spectra of the second and third models when $A_0$=100 and 200 μm in Fig. 9. When $A_0$ increases from 100 to 200 μm, the two passbands for curve 3 in the second model and for curve 2 in the third model both shift to high frequencyies. Particularly in the first cell, the passbands for curve 3 in the second model cover 0-2500 Hz instead of 0-1700 Hz, and passbands for curve 2 in the second model cover 0-2400 Hz instead of 0-1800 Hz. Therefore, curves 3 and 2 in the second and third models do exist, and they do appear only near the incident source as demonstrated in section 5.1.

In light of the the conclusions reached in sections 4.1 and 5.1, the evolution of the band structures undergoes two stages: (1) shortening and merging ($A_0$ < 20 um) and (2) shifting and response disappearnce ($A_0$ > 20 μm).

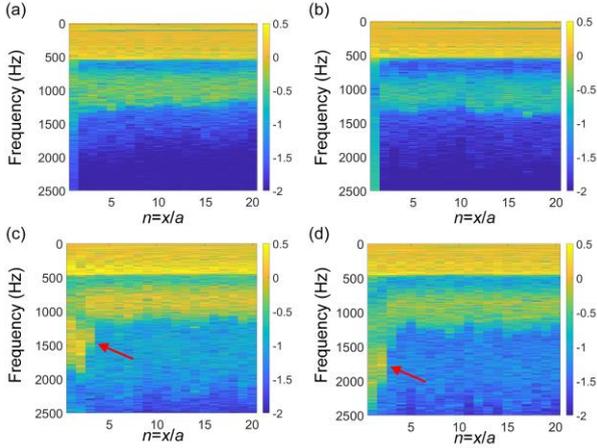

FIG. 9. Normalized amplitude Spectra of the n$^{th}$ metacell under a chirp wave excitation of 0-2500 Hz. (a), (b) Results for the second model. (c), (d) Results for the third model. (a), (c) $A_0$=100 μm. (b), (d) $A_0$=200 μm.

In the shortening and merging stage, curves 2, 3 and 4 follow the same evolution as those of curves 2,3 and 4 in section 4.1. The "disappearance of curve 3" catches up with "the arrival of curve 2", but at different frequencies. This difference shows that there is a competing mechanism between the two nonlinearities: $k_N(A_2 - A_0)^3$ and $k_N(A_2 - A_1)^3$, i.e., depending on their respective dominance nonlinearity levels.

In the shifting and response disappearnce stage, as shown in Fig. 8(b-g), for mode 1 on curve 1, from weak strong nonlinearity, $A_1/A_0$ decreases from 78.4 to 2.6, and $A_2/A_0$ from 8.6 to 1.9, i.e., both approach to the state $A_1=A_2=A_0$. This means that $m_0$, $m_1$ and $m_2$ can be considered as a merged oscillator with a total effective mass $m_0 + m_1 + m_2$. Therefore, the mechanism for the first bandgap evolves from the synchronous resonance of $m_1$ and $m_2$ to the Bragg scattering between the merged oscillator. Similarly, mode 2 also evolves from the local resonance of $m_2$ to Bragg scattering. Moreover, mode 3 nearly remains intact, like those discussed in section 4.1. Therefore, in this model, except for curve 5, all other modes evolve to Bragg scattering under very strong nonlinearity. The unit cell is gradually converted from a 3DoF system to a 2DoF, and finally to a 1DoF system.

We summarize the evolution processes of the dispersion curves in the three superlattice models in Fig. 10. The band degeneration in the first model features the shortening, merging of the high frequency dispersion curves alongside the broadening of the first resonant bandgap. The band degeneration in the second model mainly exhibits the dispersion curve shifting, response disappearance, alongside te narrowing-down of the first resonant bandgap. The third model combines these two features, underpined by a competing mechanism between the two nonlinear sources. At last, the unit cell in the superlattice evolves from 3DoF to 2DoF and then to 1DoF system upon strong nonlinearity, all consistent with the prediction.

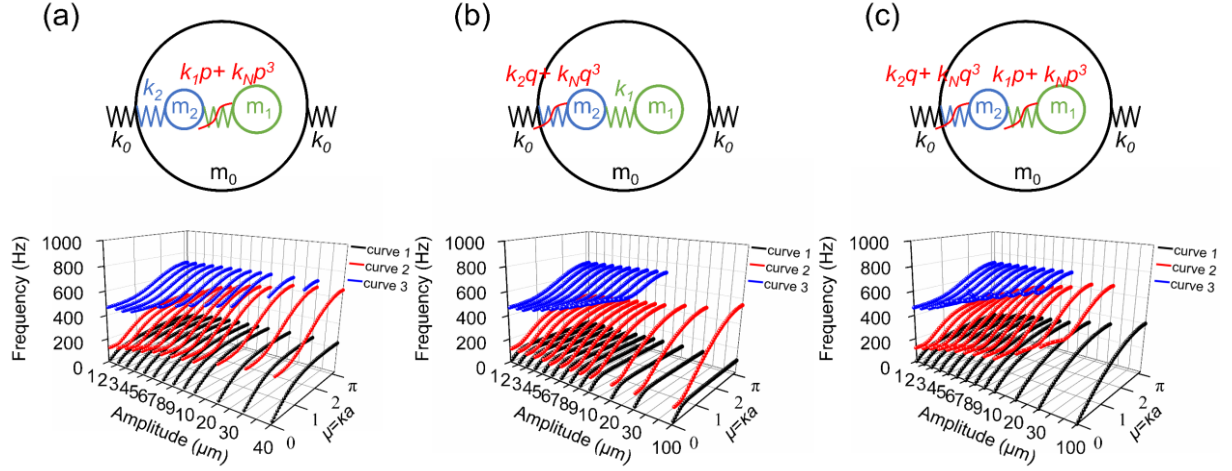

FIG. 10. The comparison of three investigated models in the band structures.

## VI. CONCLUSION

This paper studies the band degeneration and evolution in strongly nonlinear triatomic superlattices and metamaterials using analytical and numerical methods. The dispersion curves are solved with harmonic balance method and 2DFFT, which offer consisten results in the near field close to the excitation. Three typical superlattice models (the triatomic model containing two local resonators) are considered to elucidate the influences of the nonlinear coupling on the band degeneration and evolution.

We found that the dispersion curves of nonlinear superlattices may bifurcate at a wave vector for given wave amplitude, and the curves may degenerate (i.e., shorten, merge or disappear) when the excitation amplitude increases. Meanwhile, the vibration modes on the corresponding curves also vary with amplitude, typically from local resonances to Bragg scattering, thus impacting on the underlying mechanisms governing the formation of the bandgaps. A dimension reduction of the superlattice metacell is observed, which occurs due to the degeneration. Moreover, self-adaptive band structures are confirmed by 2DFFT (see Appendix).

The degeneration process depends on the arrangement of the nonlinear components. When nonlinearity appears only between two local resonators, the degeneration mainly features bifurcation, shortening and disappearance of high-frequency dispersion curves. Two locally resonant bandgaps merge into a broader resonant bandgap.

When nonlinearity only exists between a resonator and the primary oscillator, the degeneration is mainly highlighted by the shifting and response disappreance of bands. A locally resonant bandgap evolves into a Bragg bandgap alongside the narrowing-down of the remaining locally resonant bandgap. When both aforementioned nonlinarities are present, the two types of behaviors occur simultaneously: i.e. the merging of multiple bandgaps first and then the shifting and band disappearance. Furthermore, we observed that the number of dimensionality reductions is positively related to the number of nonlinear components in the system.

This work sheds light on the formation and evolution process of the band structure in strongly nonlinear superlattices and clarifies the underlying mechanisms. This is a generic yet foundamental question to be answered for studying nonlinear periodic structures, nonlinear crystals and nonlinear metamaterials. The results are therefore of interest to the broad community of nonlinear physics.

## ACKNOWLEDGMENTS

This research was funded by the Research Grant Council of the Hong Kong SAR (PolyU 152023/20E). Xin Fang is supported by the National Natural Science Foundation of China (Projects No. 12002371) and the Hong Kong Scholars Program.

## Appendix: Adaptive broadening band-gap effects

We take note that the band structures observed and reported in the main text do not consider the influence of the propagation distance, and the numerical dispersion curves are calculated for the first 50 metacells. Fang *et al*. [35] discovered and experimentally demonstrated that the band structure may adaptively vary with the propagation distance/time, i.e., self-adaptive band structure. In the present case, self-adaptive properties are shown in Fig. A1 through examining the band structures at different positions/segments. Fig. A1 (a) shows the band structure of the first 50 cells as a strong nonlinear case as demonstrated in section 4.1. When the propagation distance/time increases as shown in Fig. A1 (b), the band structure of the chain segment containing 50-100th cell exhibits moderate nonlinearity. This is due to the reduced wave amplitude as a result of the wave attenuation in the proceeding cells; meanwhile the broader range will be swept by the new bandgap, leading to increased energy reflection. Given sufficiently large propagation distance/distance increase, the band structure starts to degenerate to a weak nonlinear case, with a band structure similar to that of the linear triatomic chain shown in Fig. 1(b). Therefore, the band structure indeed varies self-adaptively as the propagation distance/time increases. In other words, the band structure of a strongly nonlinear superlattice model varies when the wave energy changes.

In addition, this self-adaptive property confirms that the Bloch theorem is no longer applicable to strongly nonlinear superlattice models, which require the consideration of spatial and temporal variation.

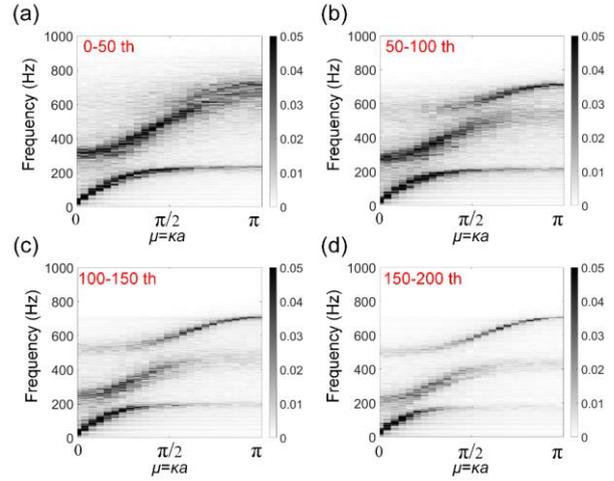

FIG. A1. Band structure of the metacells at different positions/segments with $A_0$=10 μm. (a) Metacells from 0 to 50th; (b) Metacells from 50th to 100th; (c) Metacells from 100th to 150th; (d) Metacells from 150th to 200th.

[1] Y. Rajakarunanayake, R. Miles, G. Wu, and T. McGill, Physical Review B **37**, 10212 (1988).
[2] M. Killi, S. Wu, and A. Paramekanti, Physical Review Letters **107**, 086801 (2011).
[3] L. Zou, H. C. Po, A. Vishwanath, and T. Senthil, Physical Review B **98**, 085435 (2018).
[4] S. Banerjee, B. P. Pal, and D. Roy Chowdhury, Journal of Electromagnetic Waves and Applications **34**, 1314 (2020).
[5] S. A. Cummer, J. Christensen, and A. Alù, Nature Reviews Materials **1**, 1 (2016).
[6] S. Chen, Y. Fan, Q. Fu, H. Wu, Y. Jin, J. Zheng, and F. Zhang, Applied Sciences **8**, 1480 (2018).
[7] J. Liu, H. Guo, and T. Wang, Crystals **10**, 305 (2020).
[8] K. Fan and W. J. Padilla, Materials Today **18**, 39 (2015).
[9] R. Ghaffarivardavagh, J. Nikolajczyk, S. Anderson, and X. Zhang, Physical Review B **99**, 024302 (2019).
[10] L. Wu, Y. Wang, K. Chuang, F. Wu, Q. Wang, W. Lin, and H. Jiang, Materials Today **44**, 168 (2021).
[11] V. Rothos and A. Vakakis, Wave Motion **46**, 174 (2009).
[12] S. Fiore, G. Finocchio, R. Zivieri, M. Chiappini, and F. Garesci, Applied Physics Letters **117**, 124101 (2020).
[13] A. Rose and D. R. Smith, Optical Materials Express **1**, 1232 (2011).
[14] A. Rose and D. R. Smith, Physical Review A **84**, 013823 (2011).
[15] J. Cabaret, P. Béquin, G. Theocharis, V. Andreev, V. Gusev, and V. Tournat, Physical Review Letters **115**, 054301 (2015).
[16] J. Lydon, G. Theocharis, and C. Daraio, Physical review E **91**, 023208 (2015).
[17] L. Bonanomi, G. Theocharis, and C. Daraio, Physical Review E **91**, 033208 (2015).
[18] K. L. Manktelow, M. J. Leamy, and M. Ruzzene, Wave Motion **51**, 886 (2014).
[19] R. K. Narisetti, M. Ruzzene, and M. J. Leamy, Wave Motion **49**, 394 (2012).
[20] X. Fang, J. Wen, J. Yin, and D. Yu, Aip Advances **6**, 121706 (2016).
[21] X. Fang, J. Wen, D. Yu, and J. Yin, Physical Review Applied **10**, 054049 (2018).
[22] A. Mojahed, O. V. Gendelman, and A. F. Vakakis, The

Journal of the Acoustical Society of America **146**, 826 (2019).

[23] A. Kanj, C. Wang, A. Mojahed, A. Vakakis, and S. Tawfick, AIP Advances **11**, 065328 (2021).

[24] C. Wang, A. Kanj, A. Mojahed, S. Tawfick, and A. Vakakis, Journal of Applied Physics **129**, 095105 (2021).

[25] S. Wang and V. Nesterenko, Physical Review E **91**, 062211 (2015).

[26] E. Kim, F. Li, C. Chong, G. Theocharis, J. Yang, and P. G. Kevrekidis, Physical review letters **114**, 118002 (2015).

[27] X. Fang, J. Wen, B. Bonello, J. Yin, and D. Yu, Nature communications **8**, 1 (2017).

[28] C. Daraio, V. Nesterenko, E. Herbold, and S. Jin, Physical Review E **73**, 026610 (2006).

[29] J. Cabaret, V. Tournat, and P. Béquin, Physical Review E **86**, 041305 (2012).

[30] J. M. Manimala and C. Sun, The Journal of the Acoustical Society of America **139**, 3365 (2016).

[31] R. K. Narisetti, M. J. Leamy, and M. Ruzzene, Journal of Vibration and Acoustics **132** (2010).

[32] M. Bukhari and O. Barry, Nonlinear Dynamics **99**, 1539 (2020).

[33] Z. Wu, Y. Zheng, and K. Wang, Physical Review E **97**, 022209 (2018).

[34] M. Gao, Z. Wu, and Z. Wen, Advances in Civil Engineering **2018** (2018).

[35] X. Fang, J. Wen, H. Benisty, and D. Yu, Physical Review B **101**, 104304 (2020).

[36] R. Khajehtourian and M. I. Hussein, Aip Advances **4**, 124308 (2014).